\documentclass{svjour3}                     
\smartqed  
\usepackage{graphicx}

\usepackage{natbib}

\begin{document}

\title{Mony a Mickle Maks a Muckle:\thanks{``Mony a Mickle Maks a Muckle'' is a Scottish phrase which means ``many small things add up to a big thing''. It is appropriate as this paper discusses the role of many smaller telescopes in complimenting an ELT, and secondly, as it sums up why we are interested in the minor bodies of the Solar System -- studying many small things gives us a big picture.}
}
\subtitle{Minor Body Observations with Optical Telescopes of All Sizes}


\author{Colin Snodgrass }


\institute{C. Snodgrass \at
              European Southern Observatory,
              Alonso de Cordova 3107, Vitacura, Santiago, Chile \\
               \email{csnodgra@eso.org}           
}

\date{Received: date / Accepted: date}

\maketitle

\begin{abstract}
I review the current capabilities of small, medium and large telescopes in the study of minor bodies of the Solar System (MBOSS), with the goal of identifying those areas where the next generation of Extremely Large Telescopes (ELTs) are required to progress. This also leads to a discussion of the synergies between large and small telescopes. It is clear that the new facilities that will become available in the next decades will allow us to discover smaller and more distant objects (completing size distributions) and to characterise and even resolve larger individual bodies and multiple systems, however we must also recognise that there is still much to be learned from wide surveys that require more time on more telescopes than can ever be available on ELTs. Smaller telescopes are still required to discover and characterise large samples of MBOSS.
\end{abstract}

\section{Introduction}
\label{intro}
The INAF / ESO workshop ``Future Ground based Solar System Research:
Synergies with Space Probes and Space Telescope'' aimed to discuss the potential for Solar System research provided by the next generation of telescopes (in particular the European ELT [E-ELT]) and spacecraft missions. In particular, it was designed ``to optimize scientific use and to establish synergies''. When considering the science case for an E-ELT it is important to look at the science currently achieved with smaller telescopes; in what ways do we need the larger aperture to advance? 
Also, when considering the synergies possible between ground and space based observation, it 
is also helpful to think in terms of synergies across the full range of available facilities. Here I will briefly review the capabilities of current ground based 
telescopes for studying MBOSS using optical imaging, and look to what an ELT can add to this.

\section{Scaling Laws}

An excellent discussion of the factors controlling the signal-to-noise ratio (S/N) achievable for a given magnitude with a given telescope diameter $D$ is given by \citet{AstOptics}. There are of course many parameters; the operating wavelength and bandwidth, the exposure times, the detector efficiency and pixel size, and the conditions (seeing, sky background). However, one can reasonably assume equivalent instrumentation and site conditions for the purposes of comparing telescopes of different diameters. In this case, and for a fixed wavelength, the equations of \citet{AstOptics} can be reduced to the following scaling laws, which give the limiting magnitude achievable at a given S/N:
$m_{\rm lim} \propto D$ for a given fixed value of the seeing, and
$m_{\rm lim} \propto D^2$ in the case of diffraction limited images. I use the second expression to describe the capabilities of large telescopes, since current 8-10m class and future ELTs are designed to employ adaptive optics techniques to get past the seeing limit, while for smaller telescopes the difference between the two scaling laws is small. I use the capabilities of the FORS instruments at the 8m Very Large Telescope (VLT) to form the basis of the scaling, which (from the ESO FORS web pages) have a limiting broadband imaging magnitude (S/N=5) of $m_R=26.6$ in 1 hour of observation in good, dark, conditions. Scaling this to smaller telescopes gives good agreement with, for example, EFOSC2 at the 3.6m NTT on La Silla. Scaling to a 40m ELT gives a limiting magnitude of $m_R=30$. For moving objects one often needs to consider shorter exposures to avoid background objects, and for light-curve studies (for example) one needs better S/N; I also consider the limiting magnitude in 5 minutes for S/N = 5 and 20. These give limiting magnitudes of $m_R=25.3$ and 22.2 respectively for an 8m telescope, and scale to  $m_R \approx 29$ and 26 for a 40m ELT. At the other end of the scale, these scaling laws show that for objects brighter than $m_R=18$ even the high S/N and short exposure time limit is reachable with a small ($D=1$m) telescope.

\begin{figure}
  \includegraphics[width=0.5\textwidth]{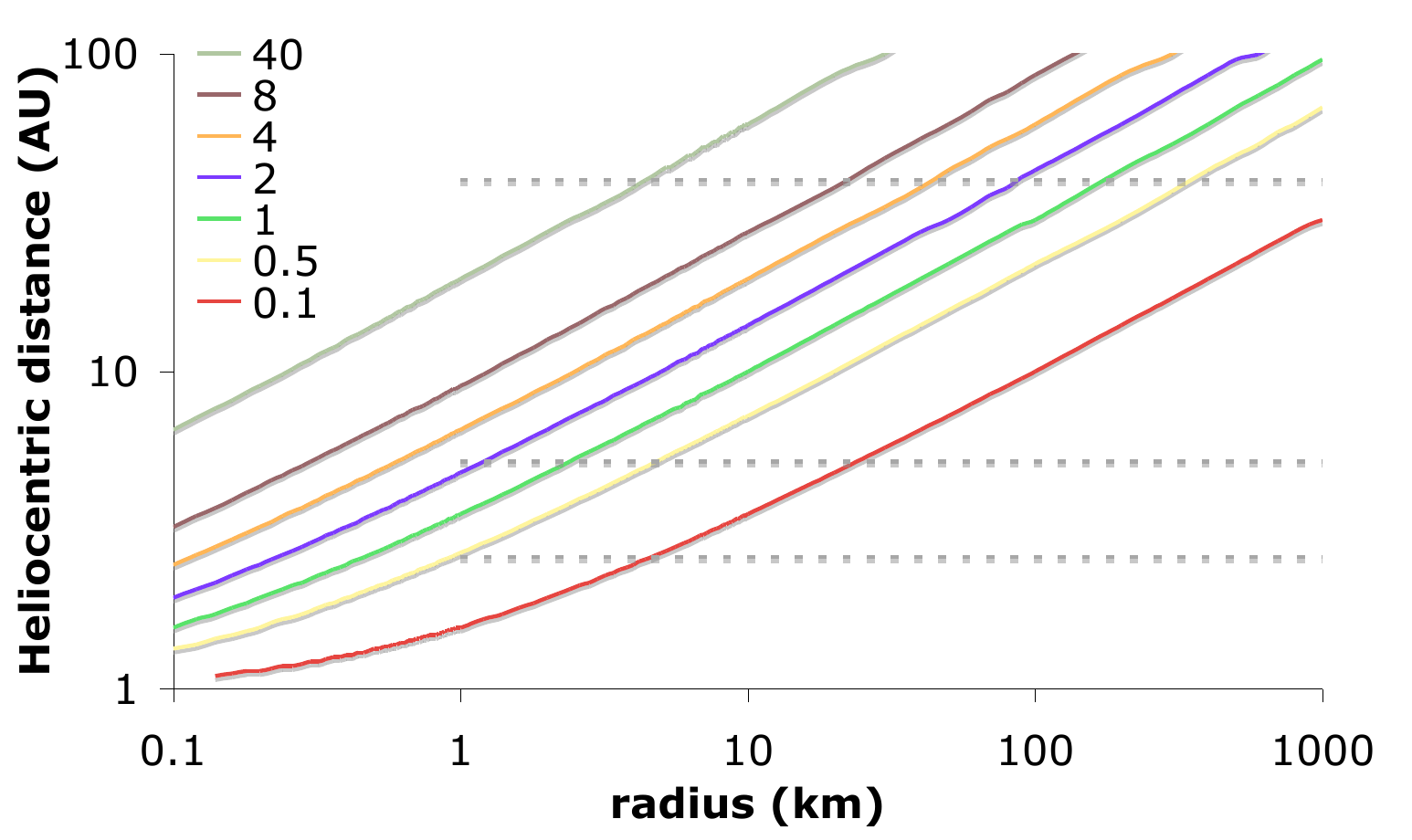}
  \includegraphics[width=0.5\textwidth]{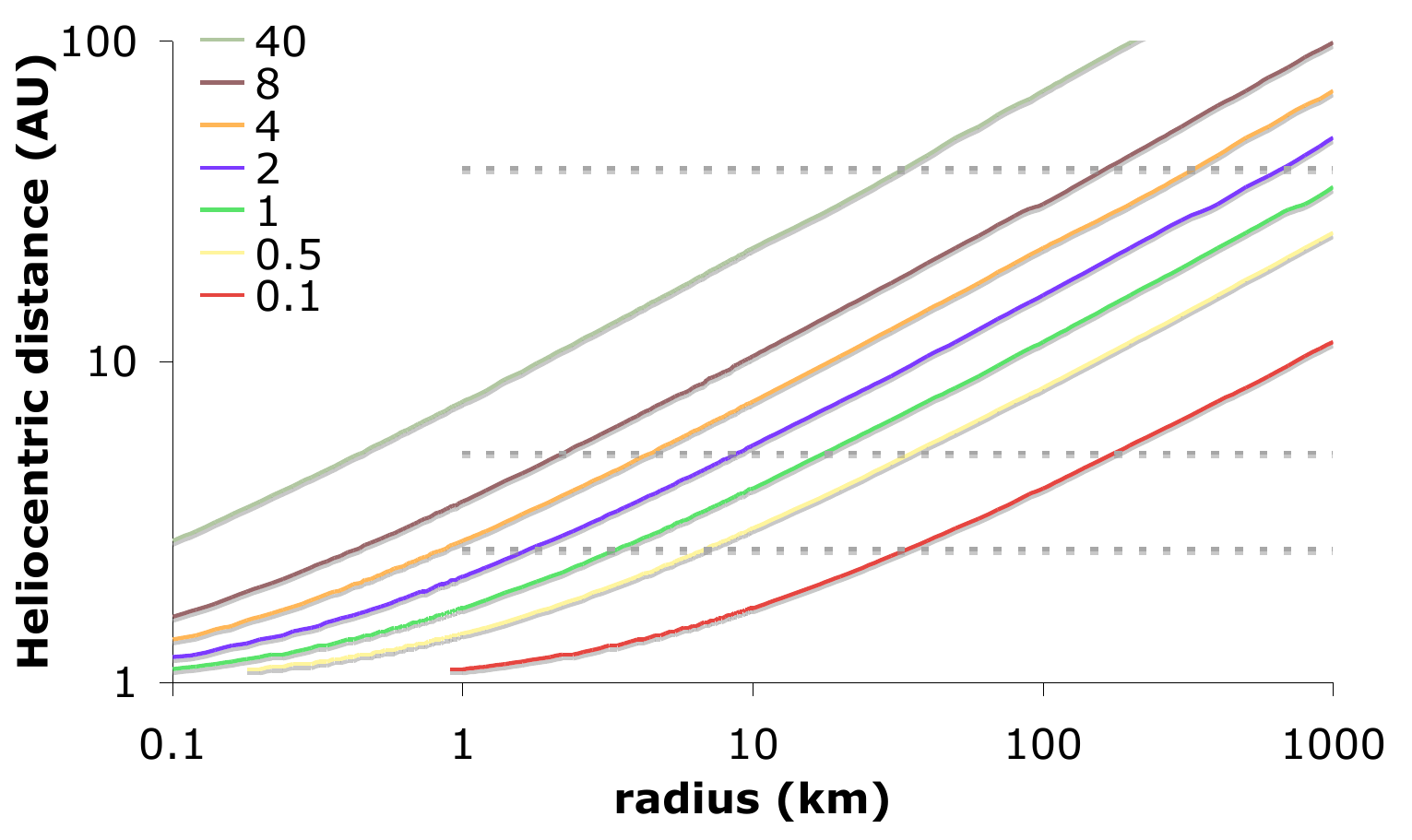}
\caption{The size of bodies that can be detected at different distances with (left) S/N=5 in a 1 hour exposure and (right) S/N=20 in a 5 minute exposure, for various telescope diameters from 0.1-40m. This assumes an inert body with a comet like albedo (4\%) at zero phase angle (and $\Delta=R_{\rm h} -1$). Many MBOSS have higher albedos (10-15\%), and consequently would be visible at slightly larger distance for the same size. Horizontal dotted lines show the distance of the Main Asteroid Belt, Jupiter (Trojans/moons/comet nuclei) and the Trans-Neptunian region (40 AU).}
\label{size_dist_fig}      
\end{figure}

For direct detection of inactive MBOSS, these magnitude limits can be expressed in terms of size and distance. In fig.~\ref{size_dist_fig} I show the detections limits for telescopes of various diameters in terms of heliocentric distance and size, for both detection in deep images (S/N=5 in 1 hour) and for study at S/N=20 in 5 minute exposures (or S/N=5 in 1 hour spectroscopy, since the limiting magnitude for medium resolution ($R \approx 2000$) spectroscopy in one hour is similar to the 5 minute S/N=20 imaging limit). 



\section{Current Science}

There are currently active areas of research across MBOSS science using telescopes of all sizes. These include discovery of objects and orbit determination, where the sensitivity follows the limits shown in fig.~\ref{size_dist_fig}, and characterising known targets. Discovery and recovery for orbit determination is done mostly by dedicated surveys (generally with small but wide field of view telescopes; see \citet{Stokes02}) but also serendipitously during other observations with telescopes of all sizes. Characterising MBOSS employ many techniques across all sizes of telescopes; I give examples using optical imaging with different sizes of existing telescopes in this section, but direct the reader to the recent books Asteroids III (2002), Comets II (2004) and The Solar System Beyond Neptune (2008) for thorough reviews of the available techniques in studying each group of objects.

Examples using the smallest telescopes ($D<1$m, e.g. the sort available to well equipped amateurs) include observation of stellar occultations by minor bodies and monitoring of the total brightness of active comets. Here the very large number of telescopes spread across the globe is critical; for occultations to have a chance of covering the ground track and to provide multiple chords \citep{Tanga+Delbo07}, and for comets to spot unexpected events such as the outburst of 17P/Holmes in October 2007 \citep{IAUC8886}. This outburst demonstrated not only the usefulness of a large network of keen amateurs, but the advantage of a network of small ($D \le 2$m), robotic, professional telescopes: We were able to follow up on the outburst with an immediate reaction, and to track the early changes in coma morphology with almost continuous coverage with world-wide spread of telescopes, without any major impact on existing programmes at these facilities. The wider field of view typical of smaller aperture telescopes was also very useful as the coma expanded, and the large amount of time that is available on such networks of small telescopes allowed us to take over 1000 images between October 2007 and January 2008 \citep{SnodgrassACM08}.
Near-Earth Objects are also suitable targets for 2m class telescopes, as even relatively small NEOs are bright enough for photometry and there are a large number of these bodies. 

Medium to large (4-8m class) current telescopes are required to observe cometary nuclei when inactive at large heliocentric distance.
We measured light-curves of comet nuclei to give shape and spin rate information, but it is difficult to get the amount of time required even on 4m class telescopes. However it is only by studying the ensemble properties of a large sample of nuclei that we are able to draw conclusions on the properties of these bodies \citep{Snodgrass06}.
For fainter nuclei 8m class telescopes are required. 
While it is generally very difficult to get time for light-curves on these, snap-shot photometry gives absolute magnitudes and colours. 

The most distant currently observed bodies are the Trans-Neptunian Objects (TNOs). The biggest and brightest are observable with small telescopes, and accessible to medium resolution spectroscopy with big telescopes, but smaller ones are more common and need 4m class telescopes for light-curves and 8m class for narrowband imaging. Even smaller bodies, down to the size of cometary nuclei, are beyond even the largest current telescopes.

\section{ELT prospects}


The limits shown in fig.~\ref{size_dist_fig} naturally lead to some projects that will require an ELT. For example, the current comet size distribution cuts off at small sizes, and whether the lack of small comets is real (i.e. a consequence of the evolution of comets) or is due to our inability to detect smaller nuclei is an important question that can be solved with a larger telescope. Likewise, the TNO size distribution will be observable down to the size of JFC nuclei, filling an important gap in the link between these populations. 

\begin{figure}
\includegraphics[width=0.5\textwidth]{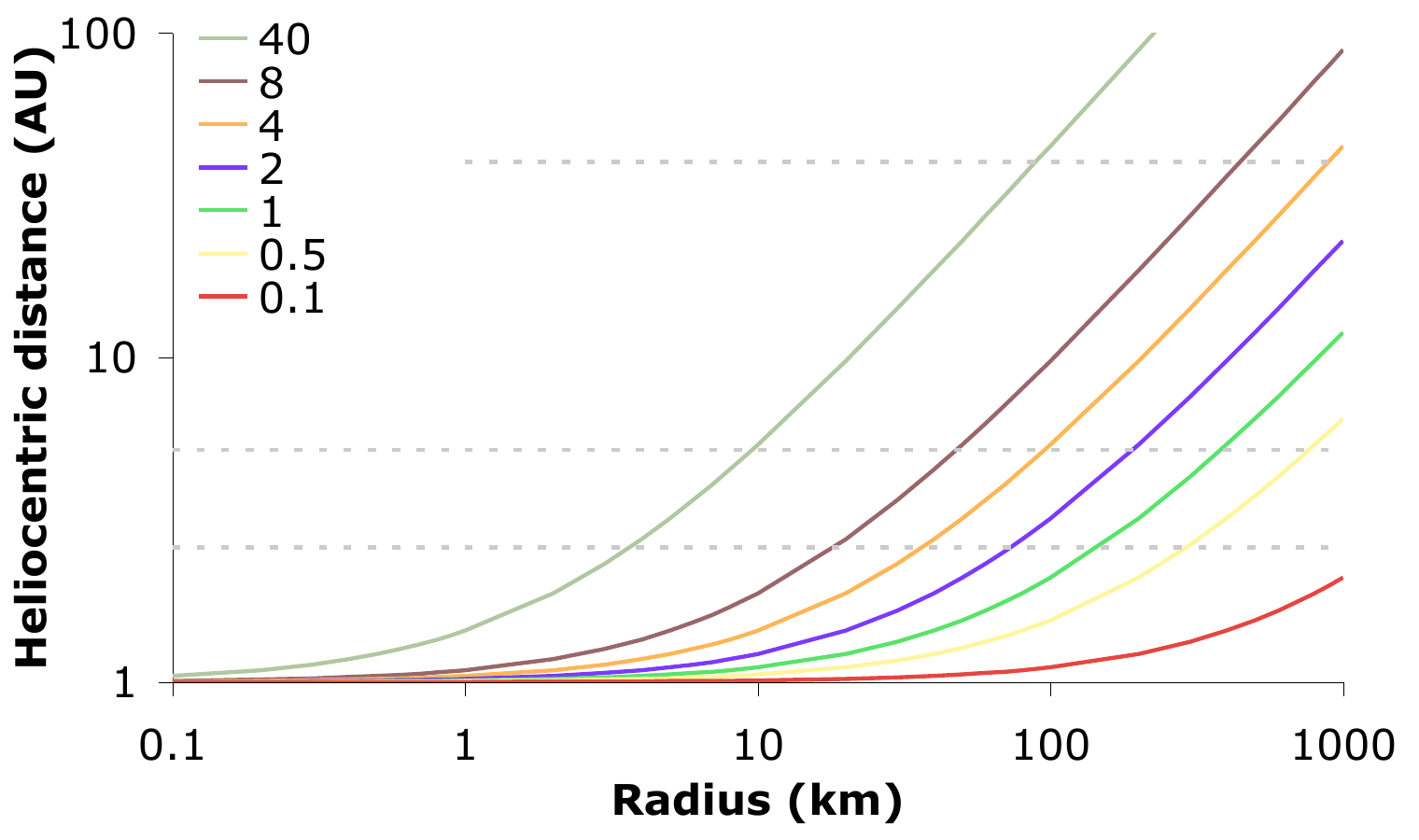}
\caption{The size of bodies that can be resolved at different distances, for various telescope diameters from 0.1-40m. This assumes adaptive optics equipped, diffraction limited telescopes. 
}
\label{resolvable}      
\end{figure}

It should be noted that size distributions based purely on optical photometry rely on assumptions about the albedo. `True' sizes are mostly found by combining this information with measurements of the thermal emission to solve for the albedo and size simultaneously \citep[e.g.][]{Seppcon}. A very few of the largest bodies are resolvable (their size can be directly measured from their apparent diameters in images) with current telescopes; the improved resolution of an ELT will allow the size of more objects to be directly measured in this way. Resolved studies also give the chance to map surface features \citep{Carry08}, pushing ground based studies into science normally only achieved by spacecraft. In fig.~\ref{resolvable} I show the size of body that can be resolved by different sized telescopes at different distances from the Sun. An ELT gives a clear advantage, with even large cometary nuclei just resolvable at the distance of Jupiter. Such resolving power will also allow separation of the bodies in binary/multiple systems, which can give a great deal of information (masses, densities of the components).

Returning to detection limits, the higher S/N side of fig.~\ref{size_dist_fig} shows that with an ELT we will be able to get light-curves and spectra (i.e. detailed characterisation of the physical properties) of objects that are currently only just detectable. Unfortunately, light-curves are likely to be too expensive in terms of telescope time to be justified.

\section{Time available}

The amount of time available on an ELT is an important issue. I briefly consider the effect of overheads: A slew time of between one minute (for modern robotic telescopes) and a few minutes is typical for a 2m telescope. 4m class telescopes tend to require of order 5 minutes, while the typical overhead to point the VLT is $\sim$ 10 minutes. The increase in overhead with diameter is partially due to the increasing size of the telescope to be moved, but mostly due to the greater complexity of the systems (closing the AO loop etc). A linear extrapolation to an ELT would suggest almost an hour per pointing, but even with a very large structure and complex full AO system this seems overly pessimistic. One could realistically expect overheads of up to 20 minutes though, meaning that a one hour OB per target would have overheads at least 50\% of the shutter time, and a program performing 5 minute observations on many independent targets would be horribly inefficient. Light-curve studies, for example, would then have to look at only one object per night. Combined with the fact that there will be only one or two massively oversubscribed ELTs worldwide (compared with $\sim$ 10 8-10m class and of order 100 smaller professional telescopes), the number of targets that can be observed with them will remain low; ELTs will give detail on single objects, not a wide survey of properties of classes of objects. 

\section{Conclusions}

There are interesting targets at the faint end of minor body distributions that will require ELTs to detect, however there remain many science projects which do not require ELTs, and indeed would be impossible with 40m telescopes. The availability of large amounts of time on small- to medium-sized telescopes remains essential for wide surveys of classes of objects, which are necessary to give context to detailed information available from ELT studies on single bodies. We need to think of `synergies' not only between future telescopes and future spacecraft, but between these new facilities and existing ones of all sizes.

\begin{acknowledgements}
I thank the referee for useful suggestions that have improved this paper.
\end{acknowledgements}


\end{document}